\pdfoutput=1
\documentclass[runningheads]{llncs}
\usepackage[T1]{fontenc}
\usepackage{graphicx}
\usepackage[hidelinks]{hyperref}

\usepackage{color}
\usepackage{bbm}
\usepackage{amsmath}
\usepackage{amsfonts}
\usepackage{makecell}
\usepackage{caption}
\usepackage{booktabs}
\usepackage{array}
\usepackage{cases}
\usepackage{threeparttable}
\usepackage{multicol}
\usepackage{enumitem}
\usepackage{marvosym}
\begin{document}
\title{DeStripe: A Self2Self Spatio-Spectral Graph Neural Network with Unfolded Hessian for Stripe Artifact Removal in Light-sheet Microscopy}
\titlerunning{DeStripe}
%
\author{Yu Liu\inst{1}\orcidID{0000-0003-2281-6791} \and
Kurt Weiss\inst{2} \and
Nassir Navab\inst{1, 3}\orcidID{0000-0002-6032-5611} \and
Carsten Marr\inst{4}\orcidID{0000-0003-2154-4552} \and
Jan Huisken\inst{2}\orcidID{0000-0001-7250-3756} \and
Tingying Peng\inst{5}\orcidID{0000-0002-7881-1749}\textsuperscript{\Letter}}
\authorrunning{Y. Liu et al.}

\institute{Technical University of Munich, Munich, Germany \and
Georg-August-University Goettingen, Goettingen, Germany \and
Johns Hopkins University, Baltimore, USA \and
Institute of AI for Health, Helmholtz Munich - German Research Center for Environmental Health, Neuherberg, Germany \and
Helmholtz AI, Helmholtz Munich - German Research Center for Environmental Health, Neuherberg, Germany\\
\email{tingying.peng@helmholtz-muenchen.de}}
\renewcommand{\thefootnote}{}
\footnotetext{This work is partly supported by the China Scholarship Council (No.202106020050).}
\maketitle              
\begin{sloppypar} 
\begin{abstract}
Light-sheet fluorescence microscopy (LSFM) is a cutting-edge volumetric imaging technique that allows for three-dimensional imaging of mesoscopic samples with decoupled illumination and detection paths. Although the selective excitation scheme of such a microscope provides intrinsic optical sectioning that minimizes out-of-focus fluorescence background and sample photodamage, it is prone to light absorption and scattering effects, which results in uneven illumination and striping artifacts in the images adversely. To tackle this issue, in this paper, we propose a blind stripe artifact removal algorithm in LSFM, called DeStripe, which combines a self-supervised spatio-spectral graph neural network with unfolded Hessian prior. Specifically, inspired by the desirable properties of Fourier transform in condensing striping information into isolated values in the frequency domain, DeStripe firstly localizes the potentially corrupted Fourier coefficients by exploiting the structural difference between unidirectional stripe artifacts and more isotropic foreground images. Affected Fourier coefficients can then be fed into a graph neural network for recovery, with a Hessian regularization unrolled to further ensure structures in the standard image space are well preserved. Since in realistic, stripe-free LSFM barely exists with a standard image acquisition protocol, DeStripe is equipped with a Self2Self denoising loss term, enabling artifact elimination without access to stripe-free ground truth images. Competitive experimental results demonstrate the efficacy of DeStripe in recovering corrupted biomarkers in LSFM with both synthetic and real stripe artifacts.


\end{abstract}
\section{Introduction}
\label{introduction}
Light-sheet Fluorescence Microscopy (LSFM) is a planar illumination technique that is revolutionizing biology by enabling rapid \emph{in toto} imaging of entire embryos or organs at subcellular resolution \cite{power2017guide,reynaud2015guide}. By illuminating the specimen perpendicular to the detection direction, LSFM excites fluorescence only in a thin slice (Fig. 1a), which allows for a higher signal-to-noise ratio and better imaging contrast \cite{mayer2018attenuation}. However, a drawback of such a lateral illumination scheme is the presence of striped artifacts along the illumination direction in the resulting image, caused by the absorption of coherent light within the sample \cite{mayer2018attenuation,ricci2021removing} (Fig. 1b). Although several optical solutions, multi-view LSFM for instance \cite{huisken2004optical}, can remove stripes in the source, they are limited by low acquisition rate and increased photobleaching, rendering them unsuitable for rapid \emph{in toto} imaging \cite{mayer2018attenuation,ricci2021removing,wei2022elimination}. Therefore, computational strategies, which attempt to remove stripe artifacts after acquisition, are highly attractive.

Inspired by the desirable properties of Fourier transform in condensing stripings into isolated values on \emph{x}-axis in Fourier space (for vertical stripes in Fig. 1b), one line of model-based destriping studies \cite{liang2016stripe,munch2009stripe} suppresses stripe noises by constructing a Fourier filter on a transformed domain, e.g., wavelet \cite{munch2009stripe}. However, filtering-based methods risk removing structural information of the sample which falls within the same filter band, resulting in image blurring negatively \cite{schwartz2019removing,chang2013robust}. On the contrary, another line of works treats the destriping issue as an ill-posed inverse problem in the standard image space, where regularizations, such as stationary prior on the stripes \cite{fehrenbach2012variational}, are commonly adopted to find the optimal solution \cite{chang2013robust,chang2016remote}. However, despite their promising abilities to preserve structural information such as sharp edges, some strict spatial constraints, e.g., low-rank assumption for the noise \cite{chang2016remote}, only hold true when the stripes cover the entire field of view, which is not the case in LSFM imaging \cite{khalilian2019strip}.

With recent advances in deep learning, emerging structural noise removal studies put image denoising tasks into a more general framework, where a mapping from a corrupted image to its noise-free counterpart is directly learned by training a generative network on a large dataset of clean/noisy images, e.g., pix2pix GAN with paired images \cite{isola2017image}, or cycleGAN on non-paired images \cite{zhu2017unpaired}. However, neither clean ground truth images \cite{weigert2018content}, nor an extensive training dataset \cite{prakash2020fully}, is easily accessible in LSFM \cite{ricci2021removing}. Encouragingly, recent developments in learning self-supervised denoising from single images, Self2Self \cite{batson2019noise2self} and Self2Void \cite{krull2019noise2void} for instance, circumvent the acquisition of clean/noisy image pairs by using the same noisy image as both input and target. For example, Self2Void proposed to randomly exclude pixels of a noise-corrupted image and optimize the denoising network only on these blind spots to prevent the model from simply learning an identical mapping \cite{krull2019noise2void}. Unfortunately, their assumption of a limited size of artifacts, which cannot span more than several connected pixels, is intrinsically not applicable to our case of striping artifacts with arbitrary shapes.

To address the aforementioned issues, in this paper, we propose a blind stripe artifact remover in LSFM, called DeStripe, by using a self-supervised spatio-spectral graph neural network with unfolded Hessian prior. The main contributions of this paper are summarized as follows:
\begin{itemize}[noitemsep,topsep=0pt]
\item[$\bullet$] DeStripe is a unified stripe artifact remover that operates in both spatial and spectral domains, enabling a complete stripe elimination by using a deep learning-parameterized Fourier filtering, while also preserving sample biological structures with an unfolded Hessian-based spatial constraint.    
\item[$\bullet$] Unlike previous convolutional image denoising networks, which adopt a U-net architecture directly in the image space to deal with artifacts spanning across multiple pixels, we formulate a graph neural network (GNN) in the spectral domain to recover stripe-affected Fourier coefficients, which is more efficient due to the isolation of stripes in Fourier space.
\item[$\bullet$] Aided by a Self2Self denoising loss formulation, DeStripe is trained completely in a self-supervised fashion, allowing blind stripe artifact removal in LSFM without the need for stripe-free LSFM images.
\end{itemize}
\begin{figure}[ht]
\centering
    \includegraphics[width=0.95\textwidth]{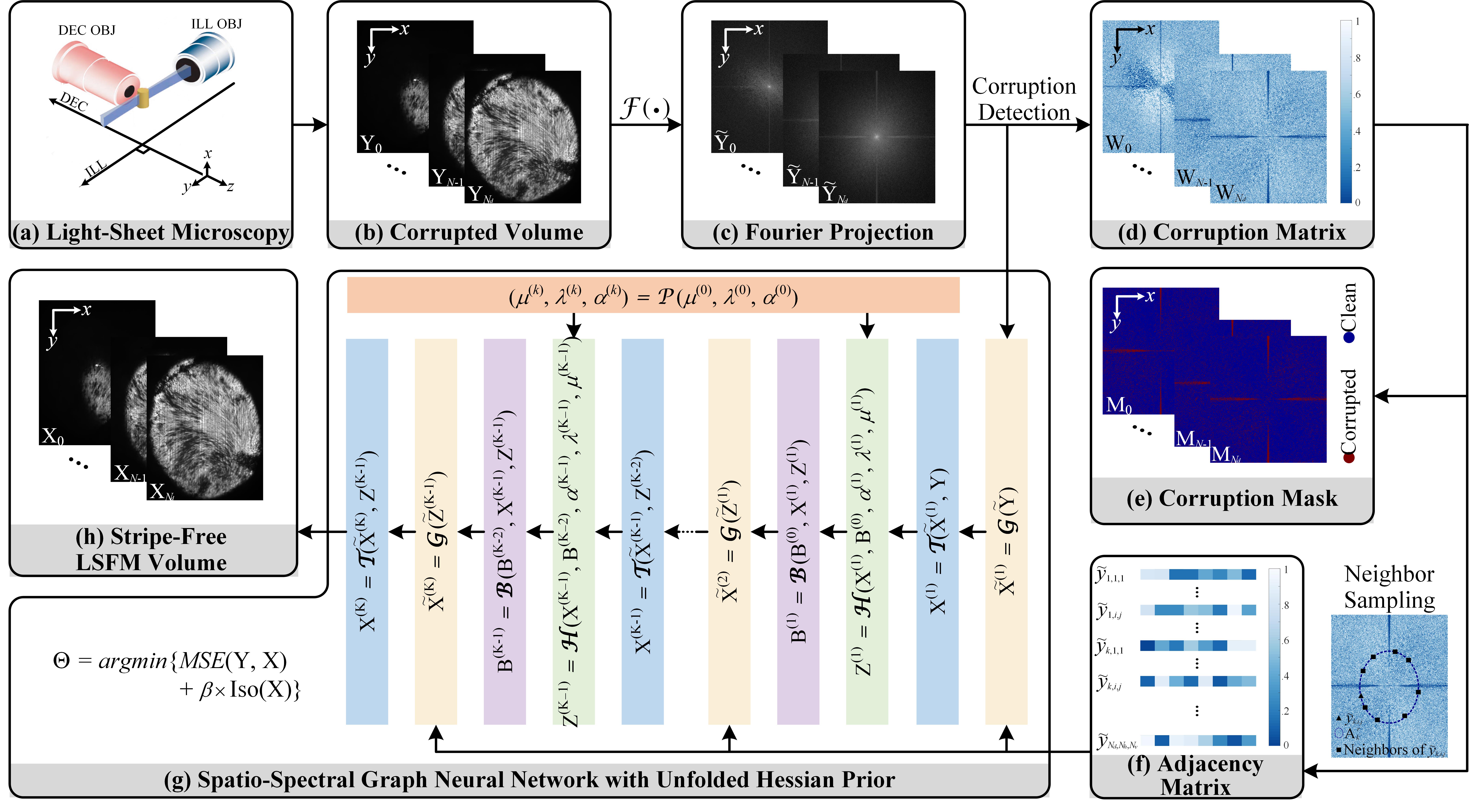}
    \caption{An overview of DeStripe (see text for explanation)}
    \label{fig:overview}
\end{figure}
\section{Methods}
\label{methods}
We illustrate DeStripe for blind stripe artifact removal in LSFM as a schematic plot in Fig. 1. First, by assuming that the Fourier projection of structured stripes is more directional than the sample itself, we locate corrupted Fourier coefficients within a wedge region in the Fourier space (Fig. 1 b-e). We then feed the affected Fourier projection into a GNN for recovery, in which the network reconstructs every noise-related Fourier coefficient based on its uncorrupted neighbors on a polar coordinate (Fig. 1f). In addition, we unfold a Hessian minimization process into our graph-based Fourier recovery network via the split Bregman algorithm (Fig. 1g), to ensure local continuity and preserve sample structure.

\subsection{Detecting of Corruption in Fourier Space}
Given a LSFM volume ${\text{Y}} \in {\mathbb{R}^{{N_d} \times {N_h} \times {N_v}}}$ with total ${N_d}$ slices of ${N_h} \times {N_v}$ images, DeStripe is to recover the underlying stripe-clean volume X from its degraded observation ${\text{Y = S}} \odot {\text{X}}$, where S is the distortion caused by stripes. In Fourier space, the spectral energy of unidirectional stripes S, which is assumed to be perpendicular to the edge in LSFM images, is highly condensed in a narrow wedge-shape frequency band perpendicular to the direction of the stripes \cite{schwartz2019removing}, whereas the underlying stripe-clean sample X has no strong direction preference in its edges (see Fig. 1c). Therefore, for every slice ${{\text{Y}}_k} \in {\mathbb{R}^{{N_h} \times {N_v}}}$, its Fourier coefficients ${\tilde y_{kij}} \in \mathbb{C}$, which fall within the same thin concentric annulus $\mathbb{A}_k^r$, mathematically follow a two-dimensional Gaussian distribution and in turn lead to the Rayleigh distribution as the amplitude distribution model \cite{khalilian2019strip}, except those stripe-corrupted ones. Therefore, a corruption matrix ${\text{W}} \in {\mathbb{R}^{{N_d} \times {N_h} \times {N_v}}}$ (Fig. 1d), whose (\emph{k}, \emph{i}, \emph{j})-th element ${w_{kij}} = S(\left\| {{{\tilde y}_{kij}}} \right\|) \in [0,1]$ indicates the degree of corresponding Fourier coefficient fulfilling the Gaussian distribution, i.e., the probability of being uncorrupted, is obtained, where $S(x) = exp( - {x^2}/2)$ is the survival function of a Rayleigh distribution \cite{khalilian2019strip}, and $\left\| {{{\tilde y}_{kij}}} \right\|$ is the magnitude of ${{{\tilde y}_{kij}}}$ after whitening. By thresholding $\text{W}$, we derive a binary corruption mask M, where $m_{i,j,k}=1$ indicates the Fourier coefficients being corrupted (Fig. 1e).

\subsection{Formulating Stripe Removal as a Deep Unfolding Framework}
\label{Formulation}
In order to recover the stripe-clean volume X from its degraded observation ${\text{Y = S}} \odot {\text{X}}$, DeStripe minimizes an energy function as follows:
\begin{equation}
\label{eq:MAP}
{\text{X}} = \mathop {argmin}\limits_{\text{X}} \left\{ {{{\left\| {{\text{Y}} - {\text{S}} \odot {\text{X}}} \right\|}^2} +  \alpha R({\text{X}}))} \right\}
\end{equation}
where the data term ${\left\| {{\text{Y}} - {\text{S}} \odot {\text{X}}} \right\|^2}$ maximizes the agreement between the prediction and input degraded image, $R({\text{X}})$ is a prior term that enforces desirable properties on the solution X, and $\alpha$ is a trade-off parameter. In DeStripe, we adopt split Bregman algorithm \cite{zhao2021sparse} to decouple the data term and prior term, resulting in three sub-problems:
\begin{subequations}
\begin{numcases}{}
{{\text{X}}^{k + 1}} = \mathop {argmin}\limits_{\text{X}} \left\{ {{{\left\| {{\text{Y}} - {\text{S}} \odot {\text{X}}} \right\|}^2} + \frac{\mu }{2}{{\left\| {{{\text{Z}}^k} - {\text{X}} - {{\text{B}}^k}} \right\|}^2}} \right\}
 \\
{{\text{Z}}^{k + 1}} = \mathop {argmin}\limits_{\text{Z}} \left\{ \alpha {R({\text{Z}}) + \frac{\mu }{2}{{\left\| {{\text{Z}} - {{\text{X}}^{k + 1}} - {{\text{B}}^k}} \right\|}^2}} \right\} \\
{{\text{B}}^{k + 1}} = {{\text{B}}^k} + {{\text{X}}^{k + 1}} - {{\text{Z}}^{k + 1}}
\end{numcases}
\end{subequations}
where $k = 1,2, \ldots ,K$ denotes the \emph{k}-th iteration, Z is introduced for splitting, B is the Bregman variable, and $\mu $ is the Lagrange multiplier. Next, in contrast to traditional model-based destriping approaches \cite{chang2013robust,khalilian2019strip,schwartz2019removing}, which derive handcrafted solutions for each sub-problem in Eq. (2), we propose to:
\begin{itemize}[noitemsep,topsep=0pt]
\item[$\bullet$] formulate a GNN-parameterized Fourier filtering to solve the data sub-problem in Eq. (2a), denoted as $\mathcal{G}( \bullet )$ yellow bar in Fig. 1g;
\item[$\bullet$] solve the prior sub-problem in Eq. (2b) with the regularizer specified as Hessian in the image space \cite{zhao2021sparse}, denoted as $\mathcal{H}( \bullet )$ green bar in Fig. 1g; 
\item[$\bullet$] adapt Bregman variable in every iteration based on Eq. (2c), denoted as $\mathcal{B}( \bullet )$ purple par in Fig. 1g;
\item[$\bullet$] inherit the hyper-parameter generator in \cite{zheng2021adaptive} as $\mathcal{P}( \bullet )$ to avoid manual parameter tuning, shown as peach bar in Fig. 1g.
\end{itemize}
\begin{figure}[ht]
\centering
    \includegraphics[width=1.0\textwidth]{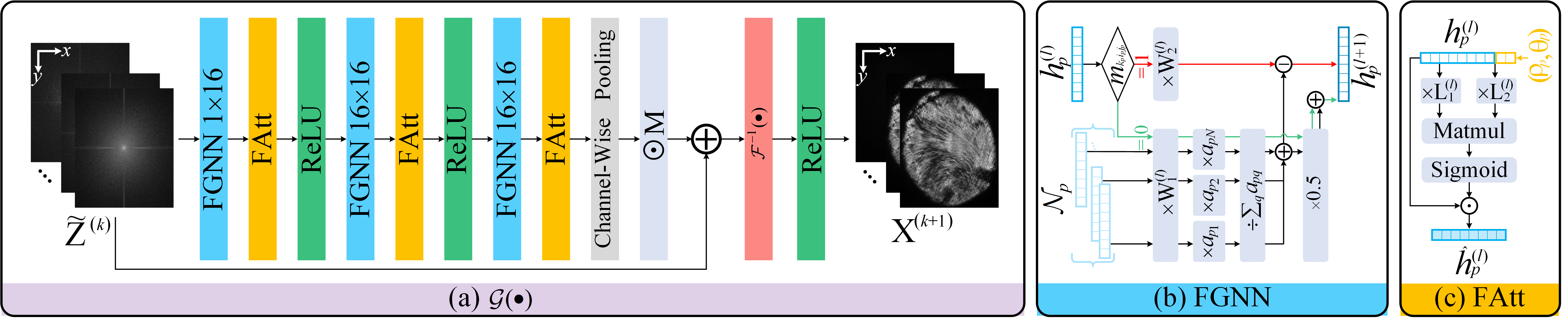}
    \caption{Structure of (a) \emph{k}-th $\mathcal{G}( \bullet )$, (b) \emph{l}-th FGNN, and (c) FAtt.}
    \label{fig:gcn}
\end{figure}

\subsection{Graph-Based Fourier Recovery Network $\mathcal{G}( \bullet )$}
\label{GNN}
Inspired by the homogeneous Fourier projection of the sample against directional one for stripings, sample-only spectral response within the corruption mask M is modeled as a combination of their uncorrupted neighbors on a polar coordinate. To this end, we adopt a GNN, which is able to vary the neighborhood size by constructing the receptive field and is shown in Fig. 2a. Specifically, we firstly reformulate Fourier projection ${\tilde{\text{Y}}} \in {\mathbb{C}^{N_d \times N_h \times N_v}}$ as a graph $\mathcal{G} = \{ \mathcal{V},{\text{H}},{\text{A}}\}$, where $\mathcal{V}$ is the vertex set with $\left| \mathcal{V} \right| = {N_d} \times {N_h} \times {N_v}$ nodes, ${\text{H}} \in {\mathbb{C}^{\left| \mathcal{V} \right| \times 1}}$ is the node attributes, whose \emph{p}-th row is the Fourier component ${\tilde y_{{k_p}{i_p}{j_p}}}$ indexed by node \emph{p}, and ${\text{A}} \in {\mathbb{R}^{|\mathcal{V}| \times |\mathcal{V}|}}$ is the adjacency matrix, whose (\emph{p}, \emph{q})-th entry indicates connection from node \emph{q} to \emph{p}. According with the isotropic hypothesis that we assume on the stripe-free X, we define neighboring connections on a polar coordinate:
\begin{equation}
\label{eq:Neighbor}
{a_{pq}} = {1_{q \in {\mathcal{N}_p}}} \times {w_{{k_q}{i_q}{j_q}}},{\text{   }}{\mathcal{N}_p}{\text{  =  }}\left\{ {q\left| {q \in \mathbb{A}_{{k_p}}^{{r_p}},} \right.{m_{{k_q}{i_q}{j_q}}}{\text{   =   0}}{\text{,}}\left| {{\mathcal{N}_p}} \right| = N} \right\}
\end{equation}
where ${\mathcal{N}_p}$ is the neighboring set of node \emph{p}, consisting of total \emph{N} uncorrupted nodes that are randomly selected from $\mathbb{A}_{{k_p}}^{{r_p}}$ of node \emph{p}. We define the proposed stripe filtering process, FGNN (Fig. 2b), as a message passing scheme on $\mathcal{G}$:
\begin{equation}
\label{eq:GCN}
h_p^{(l + 1)} = \left\{ {\begin{array}{*{20}{l}}
  {0.5 \left( {h_p^{(l)}{\text{W}}_1^{(l)} + (\sum\limits_{q \in \mathcal{N}(p)} {{a_{pq}} \times h_q^{(l)}{\text{W}}_1^{(l)}} /\sum\limits_q {{a_{pq}}} } \right),}&{{\text{  }}{m_{{k_p}{i_p}{j_p}}} = 0} \\ 
  {h_p^{(l)}{\text{W}}_2^{(l)} - (\sum\limits_{q \in \mathcal{N}(p)} {{a_{pq}} \times h_q^{(l)}{\text{W}}_1^{(l)}} )/\sum\limits_q {{a_{pq}}} ,}&{{\text{  }}{m_{{k_p}{i_p}{j_p}}} = 1} 
\end{array}} \right.
\end{equation}
where $l = 1 \ldots ,L$ is the number of layers, $h_p^{(l)} \in {\mathbb{C}^{1 \times {N_l}}}$ is the activation of node \emph{p} at the \emph{l}-th layer. Since corrupted Fourier coefficients are an accumulation of components belonging to both stripes and underlying sample, we project sample-only $h_p^{(l)}$ (${m_{{k_p}{i_p}{j_p}}} = 0$) and stripe-related $h_p^{(l)}$ (${m_{{k_p}{i_p}{j_p}}} = 1$) by ${\text{W}}_1^{(l)} \in {\mathbb{C}^{{N_l} \times {N_{(l + 1)}}}}$ and ${\text{W}}_2^{(l)} \in {\mathbb{C}^{{N_l} \times {N_{(l + 1)}}}}$ separately. 
Note that we borrow the design of complex-valued building blocks from \cite{trabelsi2018deep} for ${\text{W}}_1^{(l)}$ and ${\text{W}}_2^{(l)}$, which simulates complex arithmetic using two real-valued entities. Additionally, we insert a frequency-aware self-attention unit \cite{vaswani2017attention}, called FAtt (Fig. 2c), between every two successive FGNN, which encodes recovery importance by taking not only the Fourier coefficients but also corresponding frequencies into account. As a result, the sample-only spectral response is explicitly modeled as a weighted combination of its uncorrupted neighbors on a polar coordinate. Moreover, stripe-only Fourier projection is exclusively reserved as activation ${\text{M}} \odot {{\text{H}}^{(L+1)}}$, which can then be subtracted from the input stripe-sample mixture for striping removal.

\subsection{Unfolded Hessian Prior for Structure Preservation $\mathcal{H}( \bullet )$}
\label{Hessian}
By specifying regularizer $R({\text{X}})$ in Eq. (2b) as a Hessian prior in the image space: 
\begin{equation}
\label{eq:hessian}
\begin{split}
{R_{Hessian}}({\text{X}}) & ={\lambda _x}{\left\| {{{\text{X}}_{xx}}} \right\|_1} + {\lambda _y}{\left\| {{{\text{X}}_{yy}}} \right\|_1} + {\lambda _z}{\left\| {{{\text{X}}_{zz}}} \right\|_1} \\
& + 2\sqrt {{\lambda _x}{\lambda _y}} {\left\| {{{\text{X}}_{xy}}} \right\|_1} + 2\sqrt {{\lambda _x}{\lambda _z}} {\left\| {{{\text{X}}_{xz}}} \right\|_1} + 2\sqrt {{\lambda _y}{\lambda _z}} {\left\| {{{\text{X}}_{yz}}} \right\|_1}
\end{split}
\end{equation}
where ${\lambda _x}$, ${\lambda _y}$ and ${\lambda _z}$ are the penalty parameters of continuity along \emph{x}, \emph{y} and \emph{z} axes, respectively, ${{\text{X}}_i}$ denotes the second-order partial derivative of X in different directions. Eq. (5) then has solution as:
\begin{equation}
\label{eq:shrinkage}
{\text{Z}}_i^{k + 1} = shrink({\lambda _i} {\text{X}}_i^{k + 1}{\text{   +   B}}_i^k,\frac{\alpha }{\mu })
\end{equation}
where ${\lambda _i} = {\lambda _x},{\lambda _y},{\lambda _z},2\sqrt {{\lambda _x}{\lambda _y}} ,2\sqrt {{\lambda _x}{\lambda _z}} ,2\sqrt {{\lambda _y}{\lambda _z}}$ for $i = xx,yy,zz,xy,xz,yz$, and $shrink( \bullet )$ is the scalar shrinkage operator \cite{zhao2021sparse}.

\subsection{Self2Self Denoising Loss Formulation}
\label{loss}
We propose to train learnable parameters $ \Theta $ in DeStripe via a self-supervised denoising scheme, where training targets are still stripe-corrupted volume Y:
\begin{equation}
\label{eq:loss}
\Theta {\text{ = }}\mathop {argmin}\limits_\Theta  \left\{ {{{\left\| {{\text{Y}} - {\text{X}}} \right\|}^{\text{2}}}{\text{ + }}\beta \sum\limits_{k{\text{ = 1}}}^{{N_d}} {\sum\limits_r {\sum\limits_{\tilde x \in \mathbb{P}_k^r} {{{\left\| {\left\| {\tilde x} \right\| - \frac{1}{{\left| {\mathbb{Q}_k^r} \right|}}\sum\limits_{\tilde z \in \mathbb{Q}_k^r} {\left\| {\tilde z} \right\|} } \right\|}^2}} } } } \right\}
\end{equation}
where mean square error ${{{\left\| {{\text{Y}} - {\text{X}}} \right\|}^2}}$ is adopted to encourage the agreement between prediction X and input image Y in the image space. Particularly, the second term in Eq. (7) is to prevent the model from learning an identical mapping by quantifying isotropic properties of recovered ${{\tilde{\text{X}}}}$ in Fourier space, where $\mathbb{P}_k^r$ is the corrupted subset of $\mathbb{A}_k^r$, and $\mathbb{Q}_k^r = \left\{ {\tilde x\left| {\tilde x \in \mathbb{A}_k^r,\tilde x \notin \mathbb{P}_k^r} \right.} \right\}$.

\subsection{Competitive Methods}
We compare DeStripe to five baseline methods: (\emph{i}) wavelet-FFT \cite{munch2009stripe}: a Fourier filter-based destriping method in wavelet space; (\emph{ii}) variational stationary noise remover (VSNR) \cite{fehrenbach2012variational}: a Bayesian-based restoration framework in image space; (\emph{iii}) filling the wedge \cite{schwartz2019removing}: a total variation model-based Fourier recovery approach for stripe artifacts removal; (\emph{iv}) strip the stripes \cite{khalilian2019strip}: a Fourier reconstruction method using sparsity of the image gradient and longitudinal smoothness of the stripes for spatial constraint; (\emph{v}) SN2V \cite{broaddus2020removing}: a self-supervised deep learning network, which enables removal of structured noise by using a structured blind spots scheme; and two DeStripe variations: (\emph{vi}) DeStripe $\mathcal{G}( \bullet )$ only: constructed by removing the Hessian prior $\mathcal{H}( \bullet )$ to disable spatial constraints; (\emph{vii}) DeStripe $\mathcal{H}( \bullet )$ only: formulated by replacing $\mathcal{G}( \bullet )$ with a plain U-Net in the image space, regardless of the isolation of stripes in Fourier domain.

\begin{figure}[ht]
\centering
    \includegraphics[width=0.9\textwidth]{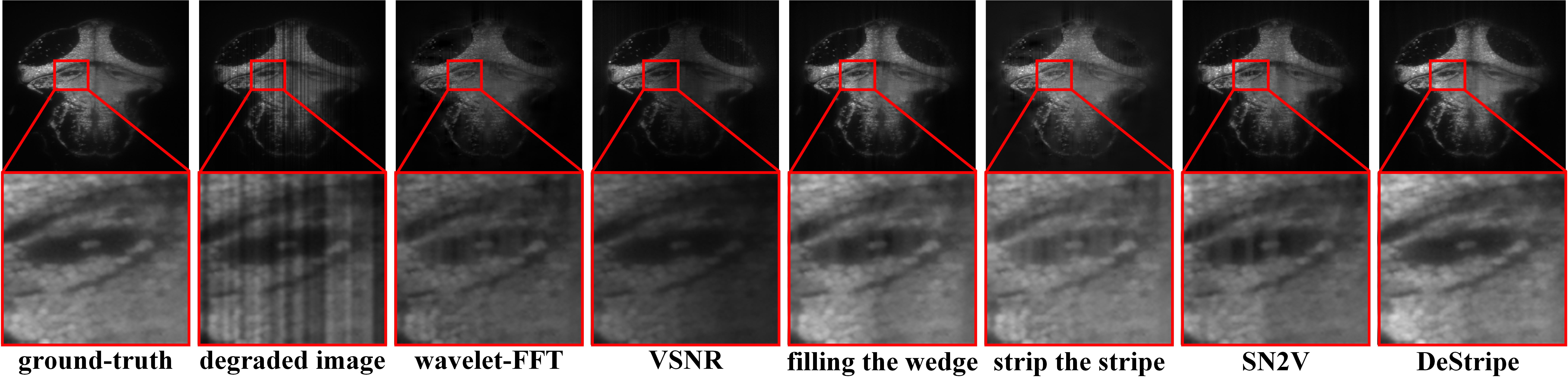}
    \caption{Visualization of stripe-removal quality with respect to ground-truth.}
    \label{fig:MainResultImg}
\end{figure}

\section{Results and Discussion}
\label{methods}
\subsection{Evaluation on LSFM images with synthetic stripe artifact}
We firstly evaluate DeStripe in removing synthetic stripes. As stated in the Introduction, there is no stripe-free LSFM with the conventional parallel light illumination. Yet special image acquisition protocol, such as diffused light-sheet microscopy, could illuminate the blind spots and alleviate stripe artifacts in the source \cite{taylor2018diffuse}. Here we take a diffused LSFM volume collected in \cite{taylor2018diffuse}, add simulated stripes following \cite{khalilian2019strip} for thirty times, perform DeStripe and other stripe removal methods, and compare the restored images to the original artifact-free ground truth. DeStripe's reconstruction achieves the best peak signal-to-noise ratio (PSNR) and structural similarity index (SSIM), far surpassing other approaches (Table.1, p<0.001 using Wilcoxon signed-rank test). It is worth noting that a SSIM of 0.98 suggests an almost flawless reconstruction by DeStripe. In comparison, wavelet-FFT, VSNR, and strip the stripes could distort the original signal gradient when removing the stripe artifacts (Fig. 3). Filling the wedge and SN2V, on the other hand, has residual stripes after correction (see enlarged image details). Only DeStripe resolves stripes without affecting original image details. Additionally, we perform an ablation study to assess individual components of DeStripe, $\mathcal{H}( \bullet )$ and $\mathcal{G}( \bullet )$. We discover that they complement one another and contribute to the overall outstanding performance.


\begin{table}[ht]
   \centering
   \begin{threeparttable}[b]
   \caption{DeStripe achieves best quantitative results on synthetic stripes.}
   \label{tab:test2}
   \centering
\begin{tabular}{c|c|c|c|c|c|c|c|c}
\toprule[2pt]
\rule{0pt}{15pt}
     & \makecell[c]{wavelet- \\ FFT\cite{munch2009stripe}}   
     & \makecell[c]{VSNR \\ \cite{fehrenbach2012variational}}          
     & \makecell[c]{filling the \\ wedge\cite{schwartz2019removing}} 
     & \makecell[c]{strip the \\ stripes\cite{khalilian2019strip}} 
     & \makecell[c]{SN2V \\ \cite{broaddus2020removing}}  
     & \makecell[c]{DeStripe \\ $\mathcal{H}( \bullet )$ only}
     & \makecell[c]{DeStripe \\ $\mathcal{G}( \bullet )$ only}
     & \textbf{DeStripe}       \\[8pt]
\hline
\rule{0pt}{15pt}
PSNR & \makecell[c]{24.25 \\ $\pm$0.77} & \makecell[c]{19.23 \\ $\pm$0.76} & \makecell[c]{31.74 \\ $\pm$1.74}    & \makecell[c]{26.13 \\ $\pm$1.56}    & \makecell[c]{20.22 \\ $\pm$0.83} & \makecell[c]{24.24 \\ $\pm$0.78}    & \makecell[c]{31.18 \\ $\pm$2.08}    & \makecell[c]{\textbf{36.34} \\ $\pm$\textbf{1.19}} \\[3pt]
\rule{0pt}{15pt}
SSIM & \makecell[c]{0.87 \\ $\pm$0.02}  & \makecell[c]{0.74 \\ $\pm$0.03}  & \makecell[c]{0.93 \\ $\pm$0.01}     & \makecell[c]{0.87 \\ $\pm$0.01}     & \makecell[c]{0.73 \\ $\pm$0.03}  & \makecell[c]{0.89 \\ $\pm$0.02}
  & \makecell[c]{0.92 \\ $\pm$0.02}.    & \makecell[c]{\textbf{0.98} \\ $\pm$\textbf{0.01}} \\[3pt]
\bottomrule[2pt]
\end{tabular}
  \end{threeparttable}
\end{table}

\begin{figure}[ht]
\centering
    \includegraphics[width=0.75\textwidth]{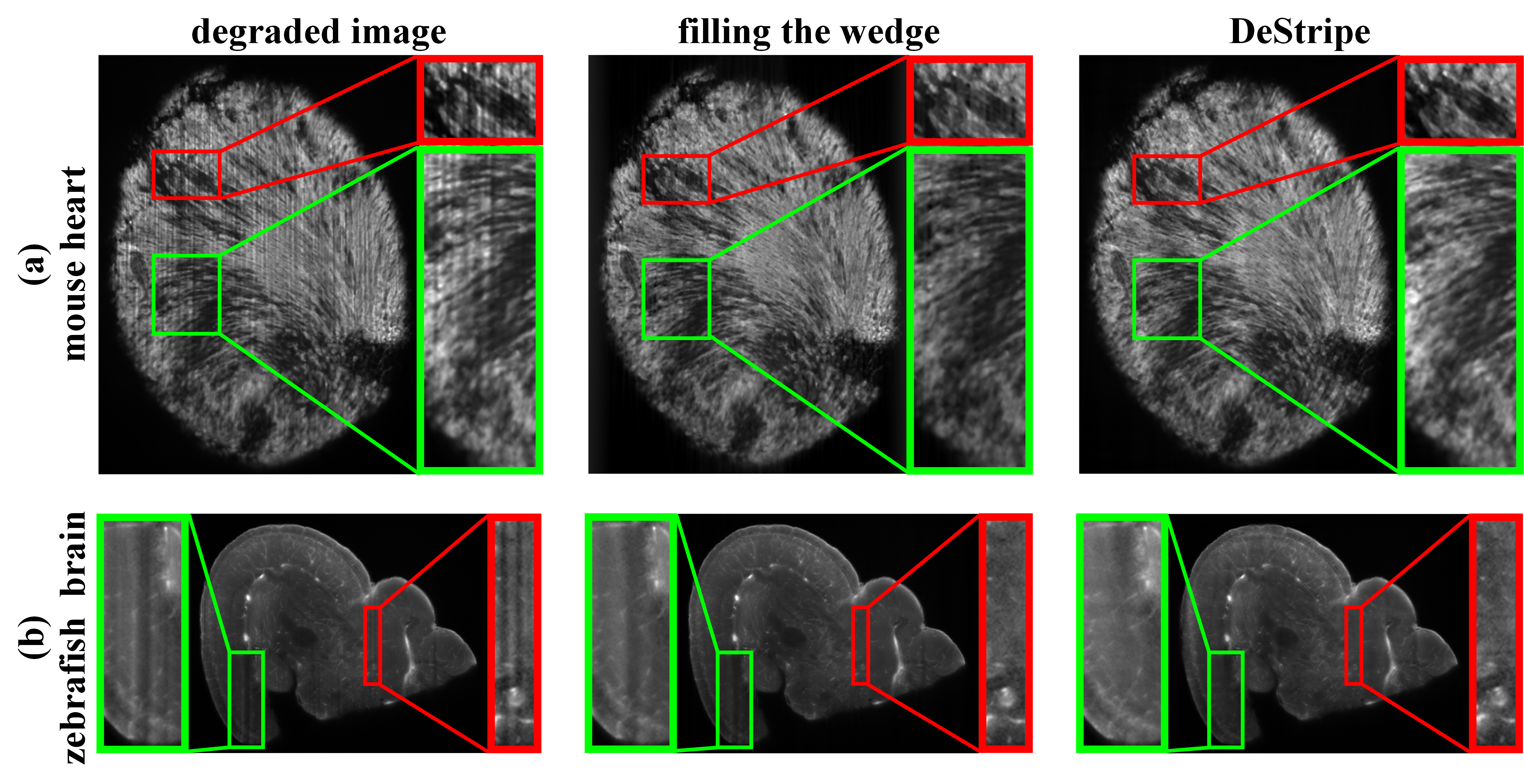}
    \caption{Visualization of stripe-removal quality in real scenario.}
    \label{fig:MainResultImg}
\end{figure}

\subsection{Evaluation on LSFM images with real stripe artifact}
DeStripe is further evaluated on real stripes in LSFM against filling the wedge \cite{schwartz2019removing}, the baseline that achieves the best performance on synthetic data. Two large sample volumes, mouse heart (100x1484x1136) and zebrafish brain (50x2169x1926), both with a resolution of 1.06 um in \emph{x}, \emph{y} and 10 um in \emph{z} axially, were optically cleared and imaged using a light sheet microscope. A multi-channel coherent laser source (Omicron Sole-6) was collimated and expanded to achieve the required light sheet size for the ca. 3 mm x 5 mm field of view and 15 um sheet waist for optical sectioning (see \cite{Huisken:07} for detailed image acquisition protocol). As shown in Fig. 4a, although filling the wedge can erase most of the quasi-periodic stripes, it also distorts the original image, e.g., causing a fake signal to appear in the formerly dark region (red box). Moreover, filling the wedge cannot resolve aperiodic thick stripes (residual stripes remain after correction, Fig. 4b). DeStripe, on the other hand, resolves both types of stripe while preserving the original image details.

\section{Conclusion}
In this paper, we propose DeStripe, a self-supervised spatio-spectral graph neural network with unfolded Hessian prior, to remove stripe artifacts in light-sheet fluorescence microscopy images. DeStripe is trained completely in a self-supervised manner, with the stripe-corrupted image serving both network input and target, obviating the need for a stripe-free LSFM for network training. Furthermore, by combing data-driven Fourier filtering in soectral domain with a Hessian-based spatial constraint, DeStripe can localize and filter isolated stripe-corrupted Fourier coefficients while better preserving sample biological structures. Both qualitative and quantitative evaluations show that DeStripe surpasses other state-of-the-art LSFM stripe removal methods by a large margin. DeStripe code will be made accessible for biologists for academic usage.


%
%
%
\bibliographystyle{splncs04}
\bibliography{bibliography}

\end{sloppypar} 
\end{document}